\documentclass[12 pt]{article}
\usepackage[ruled,vlined]{algorithm2e}
\usepackage{graphicx}
\usepackage[utf8]{inputenc}
\usepackage{amsmath,amsthm}
\usepackage{xcolor}

\SetCommentSty{mycommfont}
\SetKwInput{KwInput}{Input}                
\SetKwInput{KwOutput}{Output}              
\usepackage{mathtools}
\usepackage{amssymb}
\usepackage{latexsym}
\usepackage{indentfirst}
\usepackage{epsfig,float}
\usepackage{wrapfig,lipsum}
\usepackage{mathrsfs}
\newtheorem{cl}{Claim}
\newtheorem{thm}{Theorem}
\newtheorem{co}{Corollary}
\newtheorem{lem}{Lemma}
\newtheorem{defin}{Definition}

\title{Cosecure Domination: Hardness Results and Algorithm}
\author{Kusum\inst{1}, Arti Pandey\inst{1}}

\author{Kusum\thanks{2018maz0011@iitrpr.ac.in} \and Arti Pandey\thanks{arti@iitrpr.ac.in}}
\date{%
    Department of Mathematics, \\Indian Institute of Technology Ropar,\\ Punjab, India.\\%
   }
   
\begin{document}
\maketitle

\begin{center}
\textbf{\large{Abstract}}
\end{center}
%
%
%

For a simple graph $G=(V,E)$ without any isolated vertex, a cosecure dominating set $D$ of $G$ satisfies the following two properties (i) $S$ is a dominating set of $G$, (ii) for every vertex $v \in S$ there exists a vertex $u \in V \setminus S$ such that $uv \in E$ and $(S \setminus \{v\}) \cup \{u\}$ is a dominating set of $G$. The minimum cardinality of a cosecure dominating set of $G$ is called cosecure domination number of $G$ and is denoted by $\gamma_{cs}(G)$. The \textsc{Minimum Cosecure Domination} problem is to find a cosecure dominating set of a graph $G$ of cardinality $\gamma_{cs}(G)$. The decision version of the problem is known to be NP-complete for bipartite, planar, and split graphs. Also, it is known that the \textsc{Minimum Cosecure Domination} problem is efficiently solvable for proper interval graphs and cographs. 

In this paper, we work on various important graph classes in an effort to reduce the complexity gap of the \textsc{Minimum Cosecure Domination} problem. We show that the decision version of the problem remains NP-complete for circle graphs, doubly chordal graphs, chordal bipartite graphs, star-convex bipartite graphs and comb-convex bipartite graphs. On the positive side, we give an efficient algorithm to compute the cosecure domination number of chain graphs, which is an important subclass of bipartite graphs. In addition, we show that the problem is linear-time solvable for bounded tree-width graphs. Further, we prove that the computational complexity of this problem varies from the domination problem.

\vspace*{2mm}

%
%
%
\section{Introduction}
In this paper, $G=(V,E)$ denotes a graph without any isolated vertex, here $V$ is the set of vertices and $E$ denotes the set of edges in $G$. The graphs considered in this article are assumed to be finite, simple, undirected and without any isolated vertex. A set $D \subseteq V$ is a dominating set of graph $G$, if the closed neighbourhood of $D$ is the vertex set $V$, that is, $N[D]=V$. The domination number of a graph $G$, denoted by $\gamma(G)$, is the minimum cardinality of a dominating set of $G$. Given a graphs $G$, the \textsc{Minimum Domination} (MDS) problem is to compute a dominating set of $G$ of cardinality $\gamma(G)$. The decision version of the MDS problem is \textsc{Domination Decision} problem, notated as DM problem; takes a graph $G$ and a positive integer $k$ as an instance and asks whether there exists a dominating set of cardinality at most $k$. The \textsc{Minimum Domination} problem and many of its variations has been vastly studied in the literature and interested readers may refer to \cite{HaHeHe-20,HaHeHe-21}.

One of the important variations of domination is secure domination and this concept was first introduced by Cockayne et. al \cite{S2005} in 2005. A set $S \subseteq V$ is a \emph{secure dominating set} of $G$, if $S$ is dominating set of $G$ and for every $u \in V \setminus S$, there exists $v \in S$ such that $uv \in E$ and $(S \setminus \{u\}) \cup \{v\}$ forms a dominating set of $G$. 
The minimum cardinality of a secure dominating set of $G$ is called secure domination number of $G$ and denoted by $\gamma_s(G)$. The \textsc{Secure Domination} problem is to compute a secure dominating set of $G$ of cardinality $\gamma_s(G)$. Several researchers have contributed to the study of this problem and its many variants in
\cite{S2019,S2015,S2005,SandST2008,SDD2018}. For a detailed survey of this problem, one can refer to \cite{HaHeHe-20}.



%
Consider a situtaion in which the goal is to protect the graph by using a subset of guards and simultaneously provide a backup or substitute (non-guard) for each guard such that the resultant arrangement still protects the graph. Motivated by similar situation, another interesting variation of domination known as the cosecure domination was introduced in 2014 by Arumugam et. al \cite{CSandS2014}, which was then further studied in \cite{CS2018,K2023,CS2020,CS2021}. We can say that this variation is partly related to secure domination and that cosecure domination is, in a way, a complement to secure domination. A set $S \subseteq V$ is said to be a \emph{cosecure dominating set}, abbreviated as CSDS of $G$, if $S$ is a dominating set of $G$ and for every $u \in S$, there exists a vertex $v \in V \setminus S$ (replacement of $u$) such that $uv \in E$ and $(S \setminus \{u\}) \cup \{v\}$ is a dominating set of $G$. In this definition, we can say that $v$ \emph{$S$-replaces} $u$. A simple observation is that $V$ can never be a cosecure dominating set of $G$. It should be noted that any cosecure dominating set does not exists, if the graph have isolated vertices. Also, we remark that the cosecure domination number of a disconnected graphs $G$ is simply sum of the cosecure domination number of the connected components of $G$. So, in this paper, we will just consider only the connected graphs without any isolated vertics.

Given a graph $G$ without isolated vertex, the \textsc{Minimum Cosecure Domination} problem (MCSD problem) is an optimization problem in which we need to compute a cosecure dominating set of $G$ of cardinality $\gamma_{cs}(G)$. Given a graph $G$ without isolated vertex and a positive integer $k$, the \textsc{Cosecure Domination Decision} problem, abbreviated as CSDD problem, is to determine whether there exists a cosecure dominating set of $G$ of cardinality at most $k$. Clearly, $\gamma(G) \leq \gamma_{cs}(G)$.

The CSDD problem is known to be NP-complete for bipartite, chordal or planar graphs \cite{CSandS2014}. The bound related study on the cosecure domination number is done for some families of the graph classes \cite{CSandS2014,CS2018}. The Mycielski graphs having the cosecure domination number 2 or 3 are characterized and a sharp upper bound was given for $\gamma_{cs}(\mu(G))$, where $\mu(G)$ is the Mycielski of a graph $G$. In 2021, Zou et. al proved that $\gamma_{cs}(G)$ of a proper interval graph $G$ can be computed in linear-time \cite{CS2021}. Recently in \cite{K2023}, Kusum et al. augmented the complexity results and proved that the cosecure domination number of cographs can be determined in linear-time. They also demonstrated that the CSDD problem remains NP-complete for split graphs. In addition, they proved that the problem is APX-hard for bounded degree graphs and provided a inapproximability result for the problem. Further, they proved that the problem can be approximated within an approximation ratio of $(\Delta+1)$ for perfect graphs with maximum degree $\Delta$.

In this paper, we build on the existing research by examining the complexity status of the \textsc{Minimum Cosecure Domination} problem in many graph classes of significant importance, namely, circle graphs, doubly chordal graphs, bounded tree-width graphs, chain graphs, chordal bipartite graphs, star-convex bipartite graphs and comb-convex bipartite graphs. From the hierarchy of graph classes chordal bipartite graphs, star-convex bipartite graphs, comb-convex bipartite graphs and chain graphs are important subclasses of bipartite graphs, for which the CSDD problem is NP-complete. We reduce the gap regarding the complexity status of the problem by showing that the CSDD problem is NP-complete for chordal bipartite graphs, star-convex bipartite graphs and comb-convex bipartite graphs. We also prove that the problem remains NP-complete for doubly chordal graphs and circle graphs. On the positive side, we prove that the MCSD problem is linear-time solvable for bounded tree-width graphs and we present an efficient algorithm for computing the cosecure domination number of chain graphs. 

The structure of the rest of this paper is as follows. In section 2, we give some pertinent definitions and preliminary results. In Section 3, we demonstrate that the \textsc{Cosecure Domination Decision} problem remains NP-complete for circle graphs, chordal bipartite graphs, star-convex bipartite graphs and comb-convex bipartite graphs. In section 4, we show that the \textsc{Cosecure Domination Decision} problem is NP-complete for doubly chordal graphs. In section 5, we establish that the problem is linear-time solvable for bounded tree-width graphs. In Section 6, we give a polynomial-time algorithm for computing the cosecure domination number of chain graphs. Finally, in Section 7, we conclude the paper.

\vspace{-3mm}
\section{Preliminaries}\vspace{-2mm}

We refer to \cite{west} for graph theoretic definitions and notations. A \emph{circle graph} is a graph which is a intersection graph of chords in a circle. A graph $G=(V,E)$ is said to be a \emph{bipartite graph} if $V$ can be partitioned into $P$ and $Q$ such that for any $uv \in E$, either $u \in X$ and $v \in Y$, or $u \in Y$ and $v \in X$. Such a partition $(P,Q)$ of $V$ is said to be a \emph{bipartition} of $V$ and the sets $X$, $Y$ are called the partites of $V$. We denote a bipartite graph $G=(V,E)$ with bipartition $(P,Q)$ as $G = (P,Q,E)$ with $n_1 = \vert P\vert$ and $n_2 = \vert Q \vert$. A bipartite graph $G = (P,Q,E)$ is said to be a \emph{chordal bipartite graph}, if every cycle of length at least six has a chord. A bipartite graph $G=(P,Q,E)$ is said to be a tree-convex (star-convex or comb-convex) bipartite graph, if we can define a tree (star or comb) $T=(P,F)$ such that for every $u \in Q$, $T[N_G(u)]$ forms a connected induced subgraph of $T$ \cite{treecon}. 

A bipartite graph $G = (X, Y,E)$ is said to be a \emph{chain graph} if there exist a linearly ordering $(x_1,x_2, \ldots ,x_{n_1})$ of the vertices of the partite $X$ such that $N(x_1) \subseteq N(x_2) \subseteq \cdots \subseteq N(x_{n_1})$. If $G=(X,Y,E)$ is a chain graph, then a linear ordering, say $(y_1,y_2, \ldots ,y_{n_2})$ of the vertices of the partite $Y$ also exist such that $N(y_1) \supseteq N(y_2) \supseteq \cdots \supseteq N(y_{n_2})$. For a chain graph $G = (X, Y,E)$, a \emph{chain ordering} is an ordering $\alpha= (x_1,x_2, \ldots ,x_{n_1},y_1,y_2, \ldots ,y_{n_2})$ of $X \cup Y$ such that  $N(x_1) \subseteq N(x_2) \subseteq \cdots \subseteq N(x_{n_1})$ and $N(y_1) \supseteq N(y_2) \supseteq \cdots \supseteq N(y_{n_2})$ \cite{Chain2007}. 

Let $G=(V,E)$ is a graph. A vertex $x \in V$ is called a \emph{simplicial vertex} of $G$, if the subgraph induced on $N[x]$ is complete. A vertex $y \in N[x]$ is said to be a \emph{maximum neighbour} of $x$, if for each $z \in N[x]$, $ N[z] \subseteq  N[y]$. A vertex $x \in V$ is said to be a \emph{doubly simplicial vertex}, if $x$ is a simplicial vertex and have a maximum neighbour. A \emph{doubly perfect elimination ordering} of the vertex set $V$ of $G$, abbreviated as DPEO of $G$, is an ordering $(u_1,u_2, \ldots , u_n)$ of $V$ if for every $i \in [n]$, $u_i$ is a doubly simplicial vertex of the subgraph induced on $\{u_i,u_{i+1}, \ldots, u_n\}$ of $G$. A graph is said to be \emph{doubly chordal} if it is chordal as well as dually chordal. A characterization of doubly chordal graph is that a graph $G$ is doubly chordal if and only if $G$ has a DPEO \cite{dpeo}.

The following results are known in the literature \cite{CSandS2014}.
\vspace{-1mm}
\begin{lem}\label{CB} {\rm\cite{CSandS2014}}
For a complete bipartite graphs $G=(X,Y,E)$ with $|X| \leq |Y|$, \vspace{-4mm}
\begin{equation}
\gamma_{cs}(G)=
\begin{cases}
      |Y| & \text{if }|X| =1;\\
      2 & \text{if }|X| =2;\\
      3 & \text{if }|X| =3;\\
      4 & \text{otherwise.}
    \end{cases}
\end{equation}
\end{lem}
\vspace{-3mm}
\begin{lem}\label{support}{\rm\cite{CSandS2014}}
Let $L_u$ denote the set of pendent vertices that are adjacent to a vertex $u$ in graph $G$. If $|L_u| \geq 2$  then for every cosecure dominating set $D$ of $G$, $L_u \subseteq D$ and $u \notin D$.
\end{lem}

%

%
\section{NP-completeness results}
In this section, we study the NP-completeness of the CSDD problem in circle graphs and subclasses of bipartite graphs. The CSDD problem is known to be NP-complete for bipartite graphs and here we strengthen the complexity status of the CSDD problem by showing that it remains NP-complete for chordal bipartite graphs, star-convex bipartite graphs and comb-convex bipartite graphs which are subclasses of bipartite graphs. For that we will be using known results regarding the NP-completeness of the DM problem.

\begin{thm} {\rm \cite{dmcircle,ChB1987}}\label{dmnpc}
The DM problem is NP-complete for chordal bipartite graphs and circle graphs.
\end{thm}

\subsection{Chordal bipartite graphs and circle graphs}
In this subsection, we prove that the decision version of the \textsc{Minimum Cosecure Domination} problem is NP-complete, when restricted to chordal bipartite graphs and circle graphs. The proof of this follows by using a  polynomial-time reduction from an instance of the DM problem to an instance of the CSDD problem.

Now, we illustrate the reduction from an instance $G,k$ of the DM problem to an instance $G',k'$ of the CSDD problem. Given a graph $G=(V,E)$ with $V=\{v_i \mid 1 \leq i \leq n \}$, we construct a graph $G'=(V',E')$ from $G$ by attaching a path $(v_i,v_{i1},v_{i2})$ to each vertex $v_i \in V$, where $V'=V \cup \{v_{i1},v_{i2} \mid 1 \leq i \leq n \}$ and $E'=E \cup \{v_iv_{i1},v_iv_{i2} \mid 1 \leq i \leq n \}$. 
It is easy to see that the above defined reduction can be done in polynomial-time. The following lemma is follows from Arumungum et. al. \cite{CSandS2014}.
\begin{lem} {\rm \cite{CSandS2014}} \label{gh}
$G$ has a dominating set of cardinality at most $k$ if and only if $G'$ has a cosecure dominating set of cardinality at most $k'=k+|V(G)|$.
\end{lem}
\begin{lem}{\rm \cite{circlepen}}\label{redcircle}
Let $G$ be a circle graph and $G'$ be the graph obtained by using the above defined reduction. Then, $G'$ is also a circle graph.
\end{lem}

\begin{lem}\label{remaincb}
Let $G$ be a chordal bipartite graph and $G'$ be the graph obtained by using the above defined reduction. Then, $G'$ is also a chordal bipartite graph.
\end{lem}

The proof of following theorem directly follows from Theorem~\ref{dmnpc}, Lemma~\ref{gh}, Lemma~\ref{redcircle}, and Lemma~\ref{remaincb}.
\begin{thm} 
The CSDD problem is NP-complete for chordal bipartite graphs and circle graphs.
\end{thm}

\subsection{Star-convex bipartite graphs}
In this subsection, we prove that the decision version of the \textsc{Minimum Cosecure Domination} problem is NP-complete, when restricted to connected star-convex bipartite graphs. The proof of this follows by using a reduction from an instance of the DM problem to an instance of the CSDD problem. 
\vspace{-2mm}
\begin{thm}\label{sconvex}
The CSDD problem is NP-complete for star-convex bipartite graphs.
\end{thm}\vspace{-5mm}
\begin{proof}
Clearly, the CSDD problem is in NP for star-convex bipartite graphs. In order to prove the NP-completeness, we give a  polynomial-time reduction from the DM problem for bipartite graphs to the CSDD problem for star-convex bipartite graphs. 

Suppose that a bipartite graph $G=(X,Y, E)$ is given, where $X= \{x_i \mid 1 \leq i \leq n_1 \}$ and $Y=\{y_i \mid 1 \leq i \leq n_2 \}$. We construct a star-convex bipartite graph $G'=(X',Y', E')$ from $G$ in the following way: \vspace{-3mm}
\begin{itemize}
\item $X'=X \cup \{x,x',x_0',x_1',x_2'\}$, \vspace{-3mm}
\item $Y'=Y \cup \{y,y',y_0',y_1',y_2'\}$, and \vspace{-3mm}
\item $E'=E \cup \{xy_i, x'y_i \mid 1 \leq i \leq n_2\} \cup \{yx_i,y'x_i \mid 1 \leq i \leq n_1\} \cup \{xy_i', yx_i' \mid 1 \leq i \leq  2 \}\cup \{xy,xy',x'y,x'y',x'y_0',y'x_0'\} $. 
\end{itemize}\vspace{-2mm}

Here, $|X'|=n_1+5$, $|Y'|=n_2+5$ and $|E'|=|E|+2n_1+2n_2+10$. It is easy to see that $G'$ can be constructed from $G$ in  polynomial-time. Also, the newly constructed graph $G'$ is a star-convex bipartite graph with star $T=(X',F)$, where $F=\{xx_i \mid 1 \leq i \leq n_1\} \cup \{xx',xx_i' \mid 0 \leq i \leq 2 \}$ and $x$ is the center of the star $T$. Figure~\ref{starfig} illustrates the construction of $G'$ from $G$. 

\begin{figure}
\centering
\includegraphics[width = 11cm]{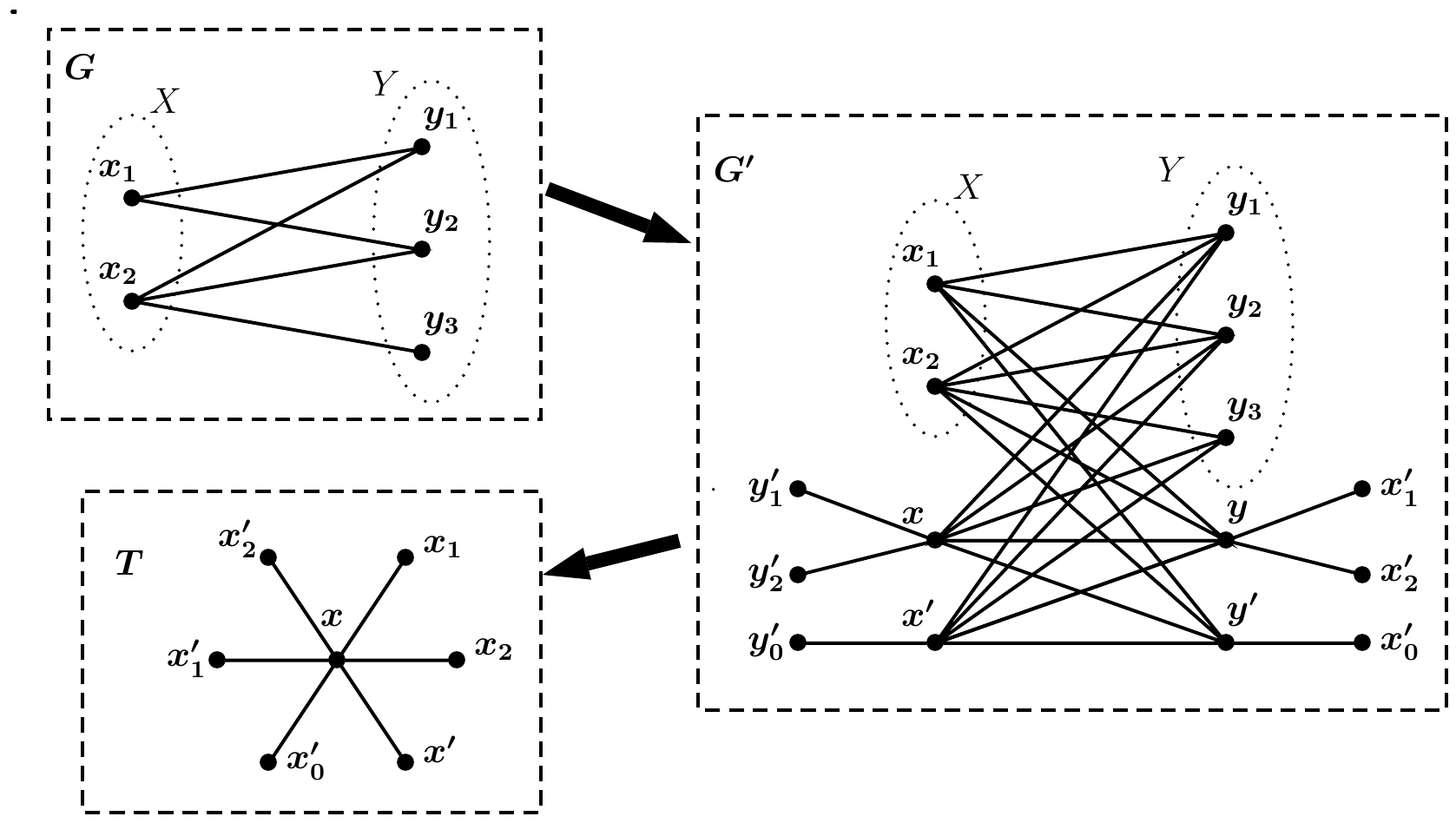}
\caption{Illustrating the construction of graph $G'$ from a graph $G$.}
\label{starfig}
\end{figure}

\begin{cl}
$G$ has a dominating set of cardinality at most $k$ if and only if $G'$ has a cosecure dominating set of cardinality at most $k+ 6$.
\end{cl}\vspace{-4mm}
\begin{proof}
Let $D$ be a dominating set of $G$ of cardinality at most $k$. Consider a set $S= D \cup \{x_i',y_i' \mid 1 \leq i \leq 2\} \cup \{x',y'\}$, where $x_i' \in X'$ and $y_i'$ for $0 \leq i \leq 2$. Clearly, $S$ is a dominating set of $G'$ and $|S|=k+6$. It is easy to see that for every vertex in $S$ there exists a replacement, as for $u \in S \cap X'\setminus \{x'\}$, $y$ is a replacement for $u$, and replacement for $x'$ is $y_0'$. Similarly, we can argue that we have replacement for each vertex $v \in S \cap Y'$. Therefore, $S$ is a cosecure dominating set of cardinality $k+6$. Hence, $G'$ has a cosecure dominating set of cardinality at most $k+ 6$.
 
Conversely, let $S$ be a cosecure dominating set of cardinality at most $k+ 6$. From Lemma~\ref{support}, it follows that $x_i',y_i' \in S$, for $1 \leq i \leq 2$ and $x,y \notin S$. Using the definition of a cosecure dominating set, it is clear that exactly one of $x'$ and $y_0'$ is in $S$. Similarly, exactly one of $y'$ and $x_0'$ is in $S$. Thus, $|S \setminus (X \cup Y) | \geq 6$. Define a set $D=S \cap (X \cup Y)$. Clearly, $|D|\le k$. Now, we claim that the set $D=S \cap (X \cup Y)$ is a dominating set of $G$. If both $x'$ and $y'$ belongs to $S$, then we are done. Note that when $x' \in S$, then $y_0'$ is the replacement for $x'$. This means that $S \cap (X \cup Y) $ dominates $X$. Similarly, we get that $S \cap (X \cup Y) $ dominates $Y$ when $y' \in S$. Therefore, we can conclude that in every possible case, $D$ form a dominating set of $G$ of cardinality at most $k$.  
\end{proof}\vspace{-3mm}
This completes the proof of the result.
\end{proof}

As tree-convex bipartite graphs is a superclass of star-conve bipartite graphs, from Theorem~\ref{sconvex} the following corollary directly follows.
\begin{co}
The CSDD problem is NP-complete for tree-convex bipartite graphs.
\end{co}
\subsection{Comb-convex bipartite graphs}
In this subsection, we prove that the decision version of the \textsc{Minimum Cosecure Domination} problem is NP-complete for comb-convex bipartite graphs. The proof of this follows by using a  polynomial-time reduction from an instance of the DM problem to an instance of the CSDD problem. 

\begin{thm}\label{cconvex}
The CSDD problem is NP-complete for comb-convex bipartite graphs.
\end{thm}\vspace{-5mm}
\begin{proof}
Clearly, the CSDD problem is in NP for comb-convex bipartite graphs. In order to prove the NP-completeness, we give a reduction from the DM problem for bipartite graphs to the CSDD problem for comb-convex bipartite graphs.

Suppose that a bipartite graph $G=(X,Y, E)$ is given, where $X= \{x_i \mid 1 \leq i \leq n_1 \}$ and $Y=\{y_i \mid 1 \leq i \leq n_2 \}$. We construct a comb-convex bipartite graph $G'=(X',Y', E')$ from $G$ in the following way: \vspace{-3mm}
\begin{itemize}
\item $X'=X \cup X^0 \cup \{x^1,x^2,x^3,a',b',c'\}$ where $X^0=\{x_i^0 \mid 1 \leq i \leq n_1 \}$, \vspace{-3mm}
\item $Y'=Y \cup \{y^1,y^2,a,b,c,d.e\} \cup \{a_i,b_i \mid 1 \leq i \leq n_1 \}$, and \vspace{-3mm}
\item $E'=E \cup \{x_iy^j, x_i^0y^j \mid 1 \leq i \leq n_1$ and $ 1 \leq j \leq 2\} \cup \{y_ix_j^0 \mid 1 \leq i \leq n_2$ and $ 1 \leq j \leq n_1\} \cup \{x^iy_j \mid 1 \leq i \leq 3$ and $ 1 \leq j \leq n_2\} \cup \{x^iy^j \mid 1 \leq i \leq 3$ and $ 1 \leq j \leq 2\} \cup \{x_i^0a_i, x_i^0 b_i \mid 1 \leq i \leq n_1\} \cup \{y^1a',y^1b',y^2c'\} \cup \{x^1a,x^1b,x^2c,x^2d,x^3e\} $. 
\end{itemize}\vspace{-2mm}

Note that $|X'|=2n_1+6$, $|Y'|=2n_1+n_2+7$ and $|E'|=|E|+6n_1+3n_2+11$. It is easy to see that 
$G'$ can be constructed from $G$ in  polynomial-time. Also, $G'$ is a comb-convex bipartite graph with comb $T=(X',F)$ where $F=\{x_i^0x_{i+1}^0 \mid 1 \leq i \leq n_1-1\} \cup \{x_ix_i^0 \mid 1 \leq i \leq n_1\} \cup \{x_{n_1}^0a',a'b',x^1x_1^0,x^1x^2,x^1x^3,x^2c' \}$ is comb with $X^0 \cup \{x^1,x^2,a'\}$ as backbone and $X \cup \{x^3,b',c'\}$ as teeth. Figure~\ref{combfig} illustrates the construction of $G'$ from $G$. 

\begin{figure}
\centering
\includegraphics[width = 11cm]{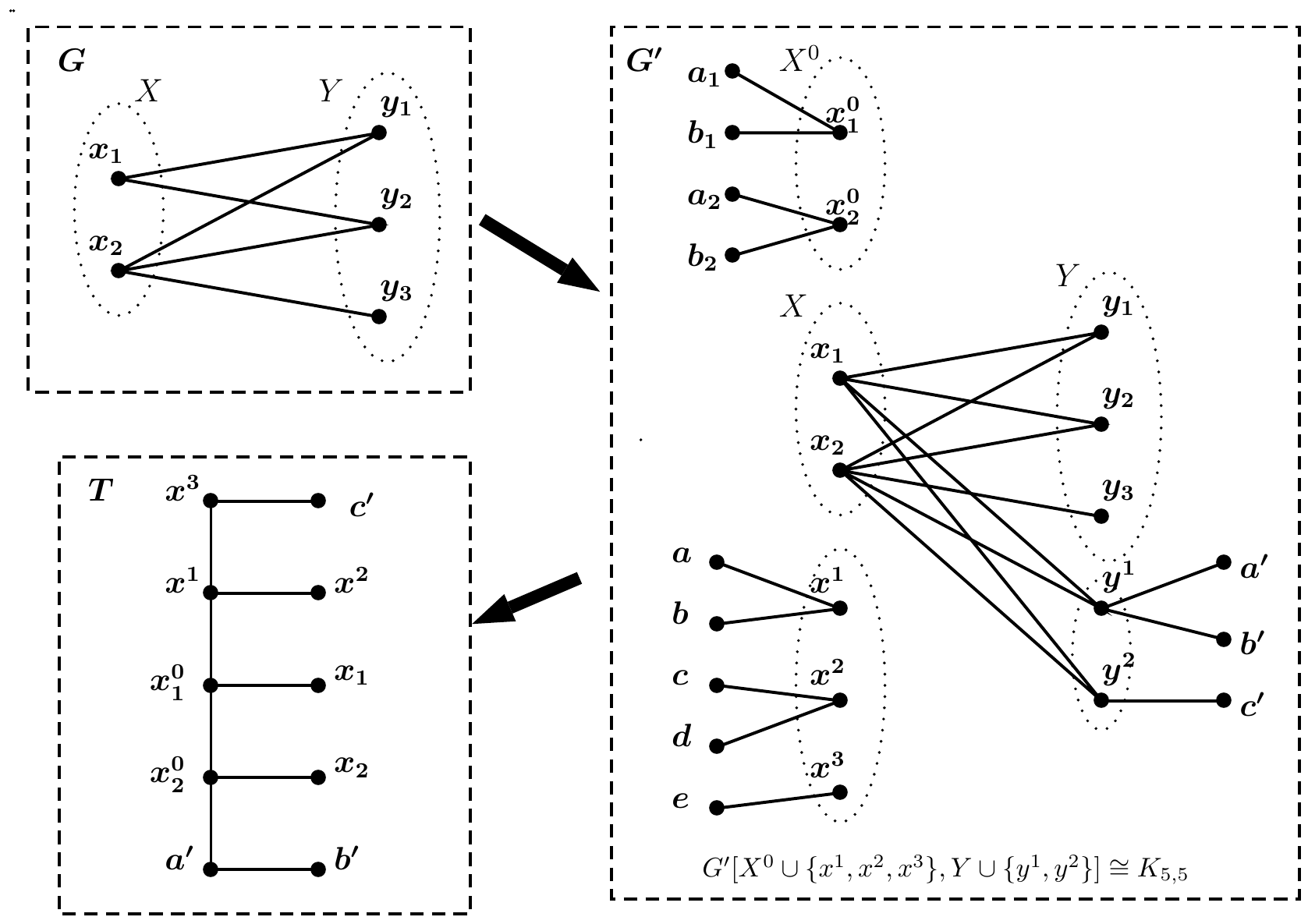}
\caption{Illustrating the construction of graph $G'$ from a graph $G$.}
\label{combfig}
\end{figure}

\begin{cl}
$G$ has a dominating set of cardinality at most $k$ if and only if $G'$ has a cosecure dominating set of cardinality at most $k+ 2(|X|+4)$.
\end{cl}\vspace{-4mm}
\begin{proof}
Let $D$ be a dominating set of $G$ of cardinality at most $k$. Consider a set $S= D \cup \{a_i,b_i \mid 1 \leq i \leq n_1 \} \cup \{a,b,c,d,a^1,b^1 ,y^2,x^3\}$. Clearly, $S$ is a dominating set of $G'$ and $|S|= k+ 2(|X|+4)$. Now, we prove that for every vertex in $S$, there exists a replacement. First, we consider the vertices in set $S \cap X'= (D\cap X) \cup \{x^3\}$ and specify a replacement for each vertex of $S\cap X'$ as follows:\vspace{-3mm}
\begin{itemize}
\item $y^1$ is a replacement for every vertex $u \in S \cap X (= D\cap X)$, \vspace{-3mm}
\item $e$ is replacement for $x^3$, and\vspace{-3mm}
\item $y^1$ is replacement for $a^1$ and $b^1$.
\end{itemize}\vspace{-2mm}

Now, we consider the vertices in set $S \cap Y'= (D\cap Y) \cup \{a,b,c,d,y^2\}$ and specify a replacement for each vertex of $S\cap Y'$ as follows:\vspace{-3mm}
\begin{itemize}
\item $x^1$ is a replacement for every vertex $u \in (S \cap Y) \cup \{a,b\}$, \vspace{-3mm}
\item $c'$ is replacement for $y^2$, \vspace{-3mm}
\item $x^2$ is replacement for $c$ and $d$,
\end{itemize}\vspace{-2mm}

Thus, for every vertex of $S$ there exists a replacement. 
Therefore, we can conclude that $S$ is a cosecure dominating set of $G'$ of cardinality $k+ 2(|X|+4)$.
 
Conversely, let $S$ be a cosecure dominating set of cardinality at most $k+ 2(|X|+4)$. From Lemma~\ref{support}, it follows that $\{ a_i,b_i \mid 1 \leq i \leq n_1 \} \cup \{ a,b,c,d,a',b' \} \subseteq S $, and $ S \cap ( \{ x^1,x^2,y^1 \} \cup X^0) = \emptyset $. Using the definition of a cosecure dominating set, it is clear that exactly one of $y^2$ and $c'$ is in $S$. Similarly, exactly one of $x^3$ and $e$ is in $S$. Thus, $|S \setminus (X \cup Y) | \geq 2(|X|+4)$. Define a set $D=S \cap (X \cup Y)$. Clearly, $|D|\le k$. Now, we claim that the set $D=S \cap (X \cup Y)$ is a dominating set of $G$. If both $c'$ and $e$ belongs to $S$, then we are done. Note that when $x^3 \in S$, then $e$ is the replacement for $x^3$. This means that $S \cap (X \cup Y) $ dominates $Y$. Similarly, we get that $S \cap (X \cup Y) $ dominates $X$ when $y' \in S$. Therefore, we can conclude that in every possible case, $D$ form a dominating set of $G$ of cardinality at most $k$.  
\end{proof}\vspace{-2mm}
This completes the proof of the result. 
\end{proof}


\section{Complexity Difference Between Domination and Cosecure Domination}
In this section, we demonstrate that the complexity of the \textsc{Minimum Domination} problem may vary from the complexity of the \textsc{Minimum Cosecure Domination} problem for some graph classes and we identify two such graph classes.

\subsection{NP-completeness of Domination for GY4-graphs}
In this subsection, we define a graph class which we call as GY4-graphs, and we prove that the MCSD problem is polynomial-time solvable for GY4-graphs, whereas the decision version of the MDS problem is NP-complete.

Let $S^4$ denote a star graph on 4 vertices. For $1 \leq i \leq n$, let $\{S^4_i \mid 1 \leq i \leq n\}$ be collection of $n$ star graphs of order 4 such that $v_i^1, v_i^2,v_i^3$ denote the pendent vertices and $v_i^4$ denote the center vertex. Now, we formally define the graph class GY4-graph as follows: 

\begin{defin}{GY4-graph}
A graph $G^Y==(V^Y,E^Y)$ is said to be a GY4-graph, if it can be constructed from a graph $G=(V,E)$ with $V=\{v_1,v_2, \ldots , v_n\}$, by making pendent vertex $v_i^1$ of a star graph $S^4_i$ adjacent to vertex $v_i\in V$, for each $1 \leq i \leq n$.
\end{defin}

Note that $|V^Y|= 4n$ and $|E^Y|= 4n+|E|$. So, $n=|V^Y|/4$. 
First, we show that the cosecure domination number can be computed in linear-time for GY4-graphs.

\begin{thm}
For a GY4-graph $G^Y=(V^Y,E^Y)$, $\gamma_{cs}(G^Y)=\frac{3}{4}|V^Y|$.
\end{thm}\vspace{-5mm}
\begin{proof}
Let $G$ be a graph with $V=\{v_1,v_2, \ldots , v_n\}$ and $G^Y$ be the GY4-graph for a graph $G$. Suppose that $D_{cs}$ is an arbitrary cosecure dominating set of $G^Y$. Using Lemma~\ref{support}, it follows that $\{ v_i^2,v_i^3 \mid 1 \leq i \leq n\} \subseteq D_{cs}$ and $v_i^1 \notin D_{cs} $. Further, observe that to dominate $v_i^1 $, at least one of $ v_{i}$ or $v_i^1$ must be there in $D_{cs}$. Collectively from above arguments, it follows that $|\{v_i,v_i^1, v_i^2,v_i^3\} \cap D_{cs}| \geq 3$, for each $1 \leq i \leq n$. Thus, $|D_{cs}| \geq 3n$. Now, using the fact that $D_{cs}$ was an arbitrary cosecure dominating set of $G^Y$ and $n=|V^Y|/4$, we have $\gamma_{cs}(G^Y) \geq \frac{3}{4}|V^Y|$. Conversely, it is easy to see that the set $D=\{ v_i^1,v_i^2 ,v_i^3 \mid 1 \leq i \leq n\}$ forms a cosecure dominating set of $G^Y$. Thus, $\gamma_{cs}(G^Y)\leq 3n$. As $n=|V^Y|/4$, we have $\gamma_{cs}(G^Y)\leq \frac{3}{4}|V^Y|$.  
\end{proof}

Next, we show that the decision version of the domination problem is NP-complete for GY-4 graphs. In order to do this, we prove that the \textsc{Minimum Domination} problem for general graph $G$ is efficiently solvable if and only if the problem is efficiently solvable for the corresponding GY4-graph $G^Y$.

\begin{lem} \label{gy4graphs}
Let $G^Y=(V^Y,E^Y)$ be a GY4-graph corresponding to a graph $G=(V,E)$ of order $n$ and $k \leq n$. Then, $G$ has a dominating set of cardinality at most $k$ if and only if $G^Y$ has a dominating set of cardinality at most $k+n$.
\end{lem}\vspace{-5mm}
\begin{proof}
Let $D^*$ be a dominating set of $G$ such that $|D^*| \leq k$. It is easy to see that $D^* \cup \{v_1^4,v_2^4, \dots , v_n^4 \}$ forms a dominating set of $G^Y$ of cardinality at most $k+n$. Conversely, let $D_Y^*$ be a dominating set of $G^Y$ of cardinality at most $k+n$. Clearly, $|D_Y \cap \{v_i^1,v_i^2,v_i^3, v_i^4 \}| \geq 1$. For each $1 \leq i \leq n$ such that $v_i^1 \in D_Y$, we can update the dominating set $D_Y$ as $D_Y\setminus \{v_i^1\} \cup \{v_i\}$. Assume that $D_Y^*$ is the updated dominating set of $G^Y$. Now, the set $D_Y^* \cap V$ forms a dominating set of $G$ of cardinality at most $k$. 
\end{proof}

As the \textsc{Domination Decision} problem is NP-complete for general graphs \cite{bipdom_npc}. Thus, the NP-completeness of the \textsc{Domination Decision} problem follows directly from Lemma~\ref{gy4graphs}.

\begin{thm} 
The DM problem is NP-complete for GY4-graphs.
\end{thm}

\subsection{NP-completeness of Cosecure Domination for Doubly Chordal graphs}
In this section, we study the NP-completeness of the CSDD problem for doubly chordal graphs. In order to prove this, we give a reduction from an instance of the \textsc{Set Cover Decision} problem to an instance of the \textsc{Cosecure Domination Decision} problem. 

Before doing that first we formally define the \textsc{Set Cover Decision} problem. Given a pair $(A,S)$ and a positive integer $k$ where $A$ is a set of $p$ elements and $S$ is a collection of $q$ subsets of $A$, the \textsc{Set Cover Decision} problem asks whether there exists a subset $S'$ of $C$ such that $\cup_{B\in S'}B=A$. The NP-completeness of the \textsc{Set Cover Decision} problem for doubly chordal graphs is already known.

\begin{thm} {\rm \cite{Setcover_npc}} \label{scdc}
The Set Cover Decision problem is NP-complete for doubly chordal graphs.
\end{thm}

\begin{thm} \label{npcdcg}
The CSDD problem is NP-complete for doubly chordal graphs.
\end{thm}\vspace{-5mm}
\begin{proof}
Clearly, the CSDD problem is in NP for doubly chordal graphs. Now, we define a reduction from the \textsc{Set Cover Decision} problem for an instance $(A,S,k)$ where $A$ is a set of $p$ elements, $S$ is a collection of $q$ subsets of $A$ and $k$ is a positive integer to an instance $(G,k')$ of the CSDD problem as follows:

Suppose that a set of elements $A=\{a_i \mid 1 \leq i \leq p\}$, collection $S=\{S_i \mid 1 \leq i \leq q\}$ of subsets of $A$ and a positive integer $k$ is given. Now we construct a graph $G=(V,E)$ in the following way:\vspace{-3mm}
\begin{itemize}
\item for each element $a_i\in A$, we take a vertex $a_i$ in $V$, \vspace{-3mm}
\item for each subset $S_i\in S$, we take a vertex $s_i$ in $V$, \vspace{-3mm}
\item $V=\{a_i \mid 1 \leq i \leq p\} \cup  \{s_i \mid 1 \leq i \leq q\} \cup  \{x_1,x_2,x_3,y_1,y_2,z_1,z_2\}$, and \vspace{-3mm}
\item $E= \{s_is_j \mid 1 \leq i<j \leq q\} \cup \{a_is_j \mid a_i\in S_j $ and $S_j \in S $, where $1 \leq i \leq p $ and $1 \leq j \leq q\}\cup \{a_ix_1,s_jx_1,s_jy_1,s_jz_1 \mid 1 \leq i \leq p $ and $1 \leq j \leq q\} \cup \{x_1x_2,x_1x_3, x_1y_1,x_1z_1, y_1z_1, y_1y_2,z_1z_2\} $. 
\end{itemize}\vspace{-2mm}

The newly constructed graph $G$ is a doubly chordal graphs with DPEO \\$\{x_2,x_3,y_2,z_2,a_1,a_2, \ldots, a_p, s_1,s_2,\ldots, s_q, y_1,z_1,x_1\}$. It is easy to see that the above construction can be done in polynomial-time. 

\begin{cl}
$(A,S)$ has a set cover of cardinality at most $k$ if and only if $G$ has a cosecure dominating set of cardinality at most $k+ 4$.
\end{cl}\vspace{-4mm}
\begin{proof}
Assume that $S'$ forms a set cover of $(A,S)$ of cardinality at most $k$. Consider $D'=\{u \mid U \in S'\}$. Define a set $D=D' \cup \{x_2,x_3,y_1,z_1\}$. It is easy to see that $D$ forms a cosecure dominating set of $G$ of cardinality at most $k+4$.

Conversely, assume that $D$ is a cosecure dominating set of $G$ of cardinality $k+4$. From Lemma~\ref{support}, it follows that $x_2,x_3 \in D$ and $x_1 \notin D$. Also, exactly one of $y_1$ and $y_2$ is in $D$, and exactly one of $z_1$ and $z_2$ is in $D$. Suppose that $I=\{a_1,a_2, \ldots, a_p\}$ and $J=\{ s_1,s_2,\ldots, s_q\}$. Above arguments implies that $|D \cap(I \cup J)| \leq k$. Now, we claim that there exists a cosecure dominating set $D'$ of $G$ such that $|D' \cap I| = \emptyset$. If $D$ satisfies $|D \cap I| = \emptyset$, then we are done. Next, assume that $D \cap I=\{u^1,u^2, \ldots, u^r\}$. If for each $u^j \in D \cap I$, there exists a vertex $w^j \in J$ such that $u^jw^j \in E$ and $w^j \notin D$, then by removing $u^j$ and adding $w^j$ in $D$, we get the required set. If for some $u^{j'} \in D \cap I$, there does not exist any vertex $w^{j'} \in J$ such that $u^jw^{j'} \in E$ and $w^{j'} \notin D$, then by simply removing such $u^{j'}$ and doing this for each such $u^{j'} \in D \cap I$, we get the required set. Thus, there exists a cosecure dominating set $D'$ of $G$ such that $|D' \cap I| = \emptyset$. Now, form a set $S'$ of subsets by including the subsets corresponding to the vertices in $D' \cap J$. As $D'$ forms a dominating set of $G$, thus the collection $S'$ of subsets forms a set cover of $(A,S)$ of cardinality at most $k$. 
\end{proof}\vspace{-2mm}
This completes the proof of the result. 
\end{proof}


\section{Bounded tree-width graphs}
In this section, we prove that the \textsc{Minimum Cosecure Domination} problem can be solved in linear-time. First, we formally define the parameter tree-width of a graph. For a graph $G=(V,E)$, its tree decomposition is a pair $(T,S)$, where $T=(U,F)$ is a tree, and $S=\{S_u \mid u \in U\}$ is a collection of subsets of $V$ such that \vspace{-3mm}
\begin{itemize}
\item $\cup_{u \in U}S_u= V$, \vspace{-3mm}
\item for each $xy \in E$, there exists $u \in U$ such that $x,y \in S_u$, and\vspace{-3mm}
\item for all $x \in V$, the vertices in the set $\{u\in U \mid x \in S_u\}$ forms a subtree of $T$.
\end{itemize}\vspace{-3mm}

The width of a tree decomposition $(T,S)$ of a graph $G$ is defined as max$\{|S_u| \mid u \in U\}-1$. The tree-width of a graph $G$ is the minimum width of any tree decomposition of $G$. A graph is said to be a bounded tree-width graph, if its tree-width is bounded. Now, we prove that the cosecure domination problem can be formulated as CMSOL.

\begin{thm}
For a graph $G=(V,E)$ and a positive integer $k$, the CSDD problem can be expressed in CMSOL.
\end{thm}\vspace{-4mm}
\begin{proof}
Let $G=(V,E)$ be a graph and $k$ be a positive integer. The CMSOL formula expressing that the existence of a dominating set $D$ of $G$ of cardinality at most $k$ is,

Dom$(D)= (D \subseteq V) \wedge (|D| \leq k) \wedge 
( (\forall x \in V)(\exists y  \in V)(( y \in D) \wedge (x \in N[y]))) $\\ 
Using the above CMSOL formula for dominating set $D$ of cardinality at most $k$, we give CMSOL formula for the cosecure dominating set of $G$ of cardinality at most $k$ as follows,

CSDM$(D)= $ Dom$(D) \wedge ((\forall x \in $ Dom$(D))(\exists y \in V \setminus D) ((y \in N(x)) \wedge $ Dom$((D\setminus \{x\}) \cup  \{y\}  ) )  )$

Hence, the result follows. 
\end{proof}

The famous  Courcelle's Theorem \cite{Cour} states that any problem which can be expressed as a CMSOL formula is solvable in linear-time for graphs having bounded tree-width. From Courcelle Theorem and above theorem, the following result directly follows.

\begin{thm}
For bounded tree-width graphs, the CSDM problem is solvable in linear-time.
\end{thm}

\vspace{-5mm}
\section{Algorithm for Chain Graphs}
In this section, we present an efficient algorithm to compute the cosecure domination number of a chain graph. Recall that a bipartite graph $G = (X, Y,E)$ is a \emph{chain graph}, if there exists a \emph{chain ordering} of $X \cup Y$, say
$\alpha= (x_1,x_2, \ldots ,x_{n_1},y_1,y_2, \ldots ,y_{n_2})$ such that  $N(x_1) \subseteq N(x_2) \subseteq \cdots \subseteq N(x_{n_1})$ and $N(y_1) \supseteq N(y_2) \supseteq \cdots \supseteq N(y_{n_2})$. Given a chain graph its chain ordering can be computed in linear-time \cite{Chain2007}. 

%
Now, we define a relation $R$ on $X$ as follows: $x_i$ and $x_j$ are related if $N(x_i)=N(x_j)$. Observe that $R$ is an equivalence relation. Assume that $X_1, X_2, \ldots, X_k$ is the partition of $X$ based on the relation $R$. Define $Y_1=N(X_1)$ and $Y_i=N(X_i) \setminus \cup_{j=1}^{i-1} N(X_j)$ for $i=2,3, \ldots k$. Then, $Y_1, Y_2, \ldots, Y_k$ forms a partition of $Y$. Such partition $X_1, X_2, \ldots, X_k,$ $Y_1, Y_2, \ldots, Y_k$ of $X \cup Y$ is called a \emph{proper ordered chain partition} of $ X \cup Y$. Note that the number of sets in the partition of $X$ (or $Y$) are $k$. Next, we remark that the set of pendent vertices of $G$ is contained in $X_1 \cup Y_k$.
%

Throughout this section, we consider a chain graph $G$ with a proper ordered chain partition $X_1, X_2, \ldots, X_k$ and $Y_1, Y_2, \ldots, Y_k$ of $X$ and $Y$, respectively. For $i \in [k]$, let $X_i=\{x_{i1}, x_{i2}, \ldots, x_{ir}\}$ and $Y_i=\{y_{i1}, y_{i2}, \ldots, y_{ir}\}$. Note that $k=1$ if and only if $G$ is a complete bipartite graph. From now onwards, we assume that $G$ is a chain graph with $k \geq 2$.

In the following lemma, we prove that if there are more than one pendent vertex from $X$ then these pendent vertices must belong to every cosecure dominating set and the corresponding support vertex does not belong to any cosecure dominating set. Note that similar result holds when there are more than one pendent vertex from $Y$. This can be generalized to the case when there are more than one pendent vertex from both $X$ and $Y$. 


\begin{lem}\label{lem2}
If there are more than one pendent from $X$, then every CSDS $S$ contains $X_1$ and does not contain $y_{11}$.
\end{lem}\vspace{-5mm}
\begin{proof}
The proof of this directly follows from Lemma~\ref{support}.
\end{proof}

Now, we assume that there are more than one pendent from $X$ in the chain graph $G$. In Lemma~\ref{sum}, we prove that the cosecure domination number of $G$ is the sum of the cosecure domination number of $G[X_1 \cup Y_1]$ and the cosecure domination number of the remaining graph. In other words, we will prove that the cosecure domination number of $G[X_1 \cup Y_1]$ and the remaining graph can be computed independently and their sum will give the cosecure domination number of $G$. Similar result follows when there are more than one pendent from $Y$.

\begin{lem}\label{sum}
Let $G$ be a chain graph such that there are more than one pendent vertex from $X$. Define $G_1=G[X_1 \cup Y_1]$, $G_2=G[ \cup_{i=2}^{k}(X_i \cup Y_i)]$. Then, $\gamma_{cs}(G)= \gamma_{cs}(G_1)+ \gamma_{cs}(G_2)$.
\end{lem}\vspace{-5mm}
\begin{proof}
Consider a chain graph $G$ such that $|X_1| \geq 2$ and $|Y_1|=1$. Assume that $G_1=G[X_1 \cup Y_1]$, $G_2=G[ \cup_{i=2}^{k}(X_i \cup Y_i)]$. 
Let $S_1$ and $S_2$ are optimal cosecure dominating sets of $G_1$ and $G_2$ respectively. Observe that $S=S_1 \cup S_2$ is a cosecure dominating set of $G$. Therefore, $\gamma_{cs}(G) \leq \gamma_{cs}(G_1)+ \gamma_{cs}(G_2)$.

Next, assume that $S$ is an optimal cosecure dominating sets of $G$. As $y_{11}$ is a support vertex and there are more than one pendent vertex adjacent to $y_{11}$. Using Lemma~\ref{lem2}, every cosecure dominating set of $G$ contains $X_1$. Therefore, $X_1 \subseteq S$. Observe that $S_1=X_1$ forms an optimal cosecure dominating set of $G_1$, so, $\gamma_{cs}(G_1)=|X_1|$. Let $S_2=S \setminus S_1$. Clearly, $S_2$ is a dominating set of $G_2$. Now, if there exists a vertex $v \in X \setminus X_1$ such that  $y_{11}$ $S-$replaces $v$. We claim that there exist $u \neq y_{11}$ such that $u$ $S-$replaces $v$. Note that $\cup_{i=2}^{j}Y_i \nsubseteq S$. To see this, let $ \cup_{i=2}^{j}Y_j \subseteq S$ then $S \setminus \{v\}$ is a CSDS of cardinality $|S|-1$ of $G$, which is a contradiction. Thus, there exists $u \in \cup_{i=2}^{j}Y_i$ such that $u \neq  y_{11}$ and $u$ $S-$replaces $v$. Observe that $u$ $S_2-$replaces $v$ as well. Thus, $S_2$ is a cosecure dominating set of $G_2$. Therefore, $\gamma_{cs}(G) \geq \gamma_{cs}(G_1)+ \gamma_{cs}(G_2)$. Hence, the result follows. 
\end{proof}

In a chain graph $G$, if there are more than one pendent vertex from both $X$ and $Y$ then using Lemma~\ref{sum}, it directly follows that the cosecure domination number of $G$ is the sum of the cosecure domination number of $G[X_1 \cup Y_1]$, the cosecure domination number of $G[X_k \cup Y_k]$ and the cosecure domination number of the remaining graph. That is, let $G_1=G[X_1 \cup Y_1]$, $G_2=G[ \cup_{i=2}^{k-1}(X_i \cup Y_i)]$ and $G_3=G[X_k \cup Y_k]$, then, $\gamma_{cs}(G)=\sum_{i=1}^{3} \gamma_{cs}(G_i)$.

Now, we consider a chain graph $G$ having $|X| \geq 4$ and $|Y| \geq 4$. In Lemma~\ref{lem3}, we give an lower bound on the cosecure domination number of $G$.  

\begin{lem}\label{lem3}
Let $G$ be a chain graph such that $|X| \geq 4$ and $|Y| \geq 4$. Then, $\gamma_{cs}(G)\geq 4$.
\end{lem}\vspace{-5mm}
\begin{proof}
Consider a chain graph $G$ such that $|X| \geq 4$ and $|Y| \geq 4$. Note that $|S| \geq 3$ as any subset of $V(G)$ of cardinality two cannot form a cosecure dominating set of $G$. Now, suppose that $S$ is a cosecure dominating set of $G$ such that $|S|=3$. Without loss of generality, assume that $|S \cap X|=2 $ and $|S \cap Y|=1 $.
Let $S \cap Y=\{y\}$ and $x \in X$ such that $x$ replaces $y$. This means that $S'=(S \setminus \{y\})\cup \{x\}$ is a dominating set of $G$. Now, let $x' \in X$ such that $x' \neq x$ and $x' \notin S$. Observe that $x'$ is not dominated by any vertex in set $S'$, which is a contradiction. Thus, there does not exist any cosecure dominating set $S$ such that $|S|=3$. Therefore, $|S|\geq 4$. Hence, the result follows.  
\end{proof}

In the next lemma, we consider the case when $G$ is a chain graph with $k=2$ and determine the cosecure domination number in all the possible cases. 

\begin{lem}\label{k2}
Let $G$ be a chain graph such that $k=2$. Then, one of the following case occurs.\vspace{-3mm}
\begin{enumerate}
\item If there does not exist any pendent vertex in $G$ and $|X|=3$ or $|Y|=3$, then $\gamma_{cs}(G)=3$, otherwise, $\gamma_{cs}(G)=4$.\vspace{-3mm}
\item If there exist more than one pendent vertex from $X$ or $Y$ or both. Define $G_1=G[X_1 \cup Y_1]$ and $G_2=G[X_2 \cup Y_2]$. Then, $\gamma_{cs}(G)= \gamma_{cs}(G_1)+ \gamma_{cs}(G_2)$.\vspace{-3mm}
\item If there exist at most one pendent vertex from $X$ and $Y$ both. If $|X|=2$ or $|Y|=2$, then $\gamma_{cs}(G)=2$. If $|X|=3$ or $|Y|=3$, then $\gamma_{cs}(G)=3$, otherwise, $\gamma_{cs}(G)=4$.
\end{enumerate}
\end{lem}\vspace{-5mm}
\begin{proof}
Consider a chain graph $G$ such that $k=2$.\vspace{-2mm}
\begin{enumerate}
\item Assume that there does not exist any pendent vertex in $G$. That is, $|X_2| \geq 2$ and $|Y_1|\geq 2$. Let $S$ be a cosecure dominating set of $G$. Note that $|S| \geq 3$. 
Now, let $|X|=3$ or $|Y|=3$. Without loss of generality, we can assume that $|X|=3$, this implies that $|X_2|= 2$ and $|X_1|=1$. Consider $S'=\{x_{11},x_{21},x_{22}\}$ then clearly $S'$ is a dominating set of $G$. As $y_{11}$ replaces every vertex of $S'$, therefore, $S'$ is a cosecure dominating set of $G$ and $|S'|=3$. Hence, $\gamma_{cs}(G)=3$. Next, we assume that $|X| \geq 4$ or $|Y| \geq 4$. Then using Lemma~\ref{lem3}, we have $|S|\geq 4$. Consider a set $S'=\{y_{11},y_{12},x_{21}, x_{22}\}$ then, clearly, $S'$ is a dominating set of $G$. As $x_{11}$ replaces both $y_{11}$ and $y_{12}$; and $y_{21}$ replaces both $x_{21}$ and $x_{22}$. Therefore, $S'$ is a cosecure dominating set of $G$ such that $|S'|=4$. Hence, $\gamma_{cs}(G)=4$.
%
\vspace{-3mm}
\item Without loss of generality, assume that there are more than one pendents in $G$ from $X$. That is, $|X_1| \geq 2$, $|Y_1|=1$ and $|X_2|\geq 2$. Let $G_1=G[X_1 \cup Y_1]$ and $G_2=G[X_2 \cup Y_2]$. Then, using Lemma~\ref{sum}, $\gamma_{cs}(G)= \gamma_{cs}(G_1)+ \gamma_{cs}(G_2)$.
\vspace{-3mm}
\item Assume that there exist at most one pendent vertex from $X$ and $Y$ both. First, let there is one pendent vertex from $X$ and $Y$ both. This implies that $|X_1|=|Y_1|=|X_2|=|Y_2|=1$ and $|X|=|Y|=2$. Then, $S=Y$ forms a cosecure dominating set of $G$. As $G$ is not a complete bipartite graph, therefore, $S$ is optimal and $\gamma_{cs}(G)=2$.

Now, consider the case when there is only one pendent vertex $u$ in $G$. Without loss of generality, let $u \in X$. Here, $|Y_1|= 1$, $|X_1|= 1$, and $|X_2|\geq 2$. If $|Y|=2$, then, $S=Y$ forms a cosecure dominating set of $G$. In fact, $S$ is an optimal cosecure dominating set of $G$ and $\gamma_{cs}(G)=2$. Now, if $|X|=3$ or $|Y|=3$. First, assume that $|X|=3$. This implies that $|X_2|= 2$. Let $S=\{x_{11},x_{21},x_{22}\}$ then, clearly, $S$ is a dominating set of $G$. As $y_{11}$ replaces every vertex of $S$, therefore, $S$ is a cosecure dominating set of $G$ and $|S|=3$. Hence, $\gamma_{cs}(G)=3$. The case when $|Y|=3$ follows similarly. Now, if $|X| \geq 4$ and $|Y| \geq 4$. Then, using Lemma~\ref{lem3}, we have $|S|\geq 4$. Let $S=\{y_{11},y_{21},x_{21}, x_{22}\}$. Clearly, $S$ is a dominating set of $G$. As $x_{11}$ replaces $y_{11}$,  $x_{23}$ replaces $y_{21}$; and $y_{22}$ replaces both $x_{21}$ and $x_{22}$. Therefore, $S$ is a cosecure dominating set of $G$, here, $|S|=4$. Hence, $\gamma_{cs}(G)=4$.

Next, assume that there are no pendent vertices from $X$ and $Y$ both, that is, $|Y_1|\geq 2$ and $|X_2|\geq 2$. Thus, $|X| \geq 3$ and $|Y| \geq 3$. If $|X|=3$ or $|Y|=3$. Without loss of generality, assume that $|X|=3$. Then, using the same arguments given in previous case we have $\gamma_{cs}(G)=3$. If $|X| \geq 4$ and $|Y| \geq 4$. Then, again using the same arguments given in previous case we have $\gamma_{cs}(G)=4$.

\end{enumerate}\vspace{-3mm}
This concludes the proof of the lemma. 
\end{proof}

From now onwards, we assume that $G$ is a connected chain graph and $k \geq 3$. In the following lemma, we will consider the case when the chain graph $G$ has no pendent vertex and we give the exact value of cosecure domination number of $G$. 

\begin{lem}\label{k3wop}
If $G$ is a chain graph without any pendent vertices, then $\gamma_{cs}(G)=4$. 
\end{lem}\vspace{-5mm}
\begin{proof}
Consider a chain graph $G$ such that $|Y_1| \geq 2$ and $|X_k| \geq 2$. Let $S$ be an optimal cosecure dominating set. Note that $|S| \geq 3$. Since $|X| \geq 4$ that $|Y| \geq 4$. Thus, using Lemma~\ref{lem3}, we have $|S|\geq 4$. Now, we claim that there exists a set $S$ such that $|S|=4$. Consider a set $S=\{y_{11},y_{12},x_{k1}, x_{k2}\}$. Observe that $S'$ is a dominating set of $G$. As $x_{11}$ replaces both $y_{11}$ and $y_{12}$; and $y_{k1}$ replaces both $x_{k1}$ and $x_{k2}$. Thus, $S'$ is a cosecure dominating set of $G$, here, $|S'|=4$. Therefore, $\gamma_{cs}(G)=4$. Hence, the result follows.  
\end{proof}

Now, we assume that in the chain graph $G$, there is at most one pendent from $X$ and $Y$ both. In Lemma~\ref{k3op}, we give the exact value of the cosecure domination number of $G$ in all the possible cases. 

\begin{lem}\label{k3op}
Let $G$ be a chain graph with at most one pendent vertex from $X$ and $Y$ both. If $|X|=3$ or $|Y|=3$, then $\gamma_{cs}(G)=3$, otherwise, $\gamma_{cs}(G)=4$.
\end{lem}\vspace{-5mm}
\begin{proof}
First, consider the case when there is no pendent vertex in graph $G$. Then, Using Lemma~\ref{k3wop}, we have $\gamma_{cs}(G)=4$. Now, assume that there is one pendent vertex from $X$ and $Y$ both. This implies that $|X_1|=|Y_1|=|X_k|=|Y_k|=1$. If $|X|=3$ or $|Y|=3$. Without loss of generality, let $|X|=3$. Let $S=\{x_{11},x_{21},x_{31}\}$. Clearly, $S$ is a dominating set of $G$. As $y_{31}$ replaces $x_{31}$; $y_{11}$ replaces $x_{11}$ and $x_{21}$. Therefore, $S$ is a cosecure dominating set of $G$ and $|S|=3$. Hence, $\gamma_{cs}(G)=3$. If $|X| \geq 4$ and $|Y| \geq 4$. Then, using Lemma~\ref{lem3}, we have $|S|\geq 4$. Consider a set $S=\{y_{11},y_{21},x_{(k-1)1}, x_{k1}\}$ then $S$ is a dominating set of $G$. Note that if $k=3$ then $|X_2| \geq 2$ and $|Y_2| \geq 2$. If $k=3$ then $x_{11}$ replaces $y_{11}$, $x_{22}$ replaces $y_{21}$, $y_{22}$ replaces $x_{21}$; and $y_{31}$ replaces $x_{31}$. Now, assume that $k \geq 4$ then $x_{11}$ replaces both $y_{11}$, $x_{21}$ replaces $y_{21}$, $y_{(k-1)1}$ replaces $x_{(k-1)1}$; and $y_{k1}$ replaces $x_{k1}$. Therefore, $S$ is a cosecure dominating set of $G$ and $|S|=4$. Hence, $\gamma_{cs}(G)=4$.

Next, assume that there is only one pendent vertex $u$ in $G$. Without loss of generality, let $u \in X$. Here, $|Y_1|= 1$, $|X_1|= 1$, and $|X_k|\geq 2$. If $|Y|=3$, then, $S=\{y_{11},y_{21}, y_{31}\}$ forms a cosecure dominating set. To see this, first observe that $S$ is a dominating set of $G$. Also, as $x_{11}$ replaces $y_{11}$ and $x_{k1}$ replaces $y_{21}$ and $y_{31}$, therefore, $S$ is a cosecure dominating set of $G$ and $|S|=3$. Hence, $\gamma_{cs}(G)=3$. Now, if $|Y| \geq 4$ then, using, Lemma~\ref{lem3}, we have $|S|\geq 4$. Consider a set $S=\{y_{11},y_{21},x_{k1}, x_{k2}\}$. Clearly, $S$ is a dominating set of $G$. As $x_{11}$ replaces both $y_{11}$, $x_{21}$ replaces both $y_{21}$; and $y_{k1}$ replaces both $x_{k1}$ and $x_{k2}$, thus, $S$ is a cosecure dominating set of $G$ and $|S|=4$. Therefore, $\gamma_{cs}(G)=4$. Hence, this completes the proof of the result. 
\end{proof}

Finally, we assume that $G$ is a chain graph such that there are at least two pendent from $X$ or $Y$ or both. In Lemma~\ref{k3mop}, we give an expression to determine the value of the cosecure domination number of $G$ in every possible case.

\begin{lem}\label{k3mop}
Let $G$ be a chain graph with $k\geq 3$. Then,\vspace{-3mm}
\begin{enumerate}
\item If there exist more than one pendent vertex from $X$ and at most one pendent from $Y$. Define $G'=G[\cup_{i=2}^{k}(X_i \cup Y_i)]$. Then, $\gamma_{cs}(G)=|X_1|+\gamma_{cs}(G')$.\vspace{-3mm}
\item If there exist more than one pendent vertex from $Y$ and at most one pendent from $X$. Define $G'=G[\cup_{i=1}^{k-1}(X_i \cup Y_i)]$. Then, $\gamma_{cs}(G)=|Y_k|+\gamma_{cs}(G')$.\vspace{-3mm}
\item If there exist more than one pendent vertex from $X$ and $Y$ both. $G'=G[\cup_{i=2}^{k-1}(X_i \cup Y_i)]$. Then, $\gamma_{cs}(G)=|X_1|+|Y_k|+\gamma_{cs}(G')$.
\end{enumerate}
\end{lem}\vspace{-5mm}
\begin{proof}
Consider a chain graph $G$ such that $k\geq 3$.\vspace{-3mm}
\begin{enumerate}
\item Let $|X_1|\geq 2$ and $|Y_1|=1$. Define $G_1=G[X_1 \cup Y_1)]$ and $G'=G[\cup_{i=2}^{k}(X_i \cup Y_i)]$. Using Lemma~\ref{sum}, $\gamma_{cs}(G)= \gamma_{cs}(G_1)+ \gamma_{cs}(G')$. As $G_1$ forms a complete bipartite graph, so, using Lemma~\ref{CB} it follows that $\gamma_{cs}(G_1)=|X_1|$. Therefore, $\gamma_{cs}(G)=|X_1|+\gamma_{cs}(G')$.\vspace{-3mm}
\item Assume that $|Y_k|\geq 2$ and $|X_k|=1$. Let us define $G_1=G[X_k \cup Y_k)]$ and $G'=G[\cup_{i=1}^{k-1}(X_i \cup Y_i)]$. Then, using Lemma~\ref{sum}, $\gamma_{cs}(G)= \gamma_{cs}(G_1)+ \gamma_{cs}(G')$. Since $G_1$ is a complete bipartite graph, thus, using Lemma~\ref{CB} it follows that $\gamma_{cs}(G_1)=|Y_k|$. Therefore, $\gamma_{cs}(G)=|Y_k|+\gamma_{cs}(G')$.
\vspace{-3mm}
\item Let $|X_1|\geq 2$, $|Y_k|\geq 2$ and $|X_k|=|Y_1|=1$. 
Define $G_1=G[X_1 \cup Y_1)]$ and $G_2=G[\cup_{i=2}^{k}(X_i \cup Y_i)]$.  Since $|X_1|\geq 2$ and $|Y_1|=1$, thus, using Lemma~\ref{sum}, $\gamma_{cs}(G)= \gamma_{cs}(G_1)+ \gamma_{cs}(G_2)$. As $G_1$ forms a complete bipartite graph, so, using Lemma~\ref{CB} it follows that $\gamma_{cs}(G_1)=|X_1|$. Thus, $\gamma_{cs}(G)=|X_1|+\gamma_{cs}(G_2)$. Now, consider the chain graph $G_2$ and define $G_3=G[X_k \cup Y_k)]$ and $G'=G[\cup_{i=2}^{k-1}(X_i \cup Y_i)]$. Now, $|Y_k|\geq 2$ and $|X_k|=1$, using Lemma~\ref{sum}, $\gamma_{cs}(G_2)= \gamma_{cs}(G_3)+ \gamma_{cs}(G')$. Since $G_3$ is a complete bipartite graph, using Lemma~\ref{CB} it follows that $\gamma_{cs}(G_3)=|Y_k|$. Thus, $\gamma_{cs}(G_2)=|Y_k|+\gamma_{cs}(G')$. Therefore, $\gamma_{cs}(G)=|X_1|+\gamma_{cs}(G_2)$ implies that $\gamma_{cs}(G)=|X_1|+|Y_k|+\gamma_{cs}(G')$. 
\end{enumerate}\vspace{-3mm}
This completes the proof of the result. 
\end{proof}

Before designing our algorithm for connected chain graphs, we first give a simple algorithm, namely \textbf{CSDN$\_$CB$(G,p,q)$} that computes the cosecure domination number of a complete bipartite graph. This algorithm is designed using Lemma~\ref{CB}. The algorithm \textbf{CSDN$\_$CB$(G,p,q)$} takes a complete bipartite graph and cardinalities of the partite sets, namely $p,q$ satisfying $p\leq q$ as input and returns $\gamma_{cs}(G)$ as output.

\begin{algorithm}[]
\label{algori}
\textbf{Input:} A complete bipartite graph $G=(X,Y,E)$ with $|X| \leq |Y|$ and two integers $p,q$ where $p=|X|$ and $q=|Y|$. \\
\textbf{Output:} Cosecure domination number $\gamma_{cs}(G)$.\\
\If{$(p=1)$}{
 $\gamma_{cs}(G)=q$;
 return $\gamma_{cs}(G)$;
}

\If{$(p= 2)$}{
 $\gamma_{cs}(G)=2$;
 return $\gamma_{cs}(G)$;
}

\If{$(p= 3)$}{
 $\gamma_{cs}(G)=3$;
 return $\gamma_{cs}(G)$;
}
\If{$(p\geq 4)$}{
 $\gamma_{cs}(G)=4$;
 return $\gamma_{cs}(G)$;
}

\caption{\textbf{CSDN$\_$CB$(G,p,q)$}}
\end{algorithm}

Now, based on the above lemmas, we design a recursive algorithm, namely,\\ \textbf{CSDN$\_$Chain$(G,k)$} to find the cosecure domination number of chain graphs. The algorithm takes a connected chain graph $G=(V,E)$ with a proper ordered chain partition $X_1,X_2, \ldots, X_k$ and $Y_1,Y_2, \ldots, Y_k$ of $X$ and $Y$ as an input. While executing the algorithm, we call the algorithm \textbf{CSDN$\_$CB$(G,p,q)$} whenever we encounter a complete bipartite graph. 

Let $G$ be a connected chain graph and $X_1,X_2, \ldots, X_k$ and $Y_1,Y_2, \ldots, Y_k$ be the proper ordered chain partition of $X$ and $Y$, respectively. The case when $k=2$ works as base case of our algorithm. The correctness of the base case follows from Lemma~\ref{k2}. Then, Lemma~\ref{k3mop} helps us in designing the algorithm using the recursive approach and proves that the correctness of the algorithm. Now, we state the main result of this section. The proof of the following theorem directly follows from combining Lemma~\ref{k3wop}, Lemma~\ref{k3op} and Lemma~\ref{k3mop}. As the running time of our algorithm  \textbf{CSDN$\_$Chain$(G,k)$} is polynomial, therefore, the cosecure domination number of a connected chain graph can be computed in  polynomial-time.

\begin{algorithm}[]
\label{algo}
\small{
\textbf{Input:} A connected chain graph $G=(V,E)$ with proper ordered chain partition $X_1,X_2, \ldots, X_k$ and $Y_1,Y_2, \ldots, Y_k$ of $X$ and $Y$. \\
\textbf{Output:} Cosecure domination number of $G$, that is, $\gamma_{cs}(G)$.\\

\If{$(k=2)$}{
 	\If{$(|Y_1| >1$ and $|X_2|>1)$}{
 	$X=X_1 \cup X_2$, $Y=Y_1 \cup Y_2$;\\
 	\If{$(|X| =3$ or $|Y|=3)$}{
		$\gamma_{cs}(G)=3$;
		}
	\Else{ 
		$\gamma_{cs}(G)=4$;
		}
	}
	\ElseIf{$((|X_1| >1$ and $|Y_1|=1)$ or $(|X_2| =1$ and $|Y_2|>1))$}{
	Let $G_1=G[X_1 \cup Y_1]$ and $G_2=G[X_2 \cup Y_2]$;\\
	Let $p_1=$min$\{|X_1|,|Y_1|\}$, $q_1=$max$\{|X_1|,|Y_1|\}$, $p_2=$min$\{|X_2|,|Y_2|\}$ and $q_2=$max$\{|X_2|,|Y_2|\}$;\\
	$\gamma_{cs}(G)=$\textbf{CSDN$\_$CB}$(G_1,p_1,q_1)+$\textbf{CSDN$\_$CB}$(G_2,p_2,q_2)$;
	}
	\ElseIf{$((|X_1| =|Y_1|=1)$ or $(|X_2| =|Y_2|=1))$}{	
	\If{$(|X| =2$ or $|Y|=2)$}{
		$\gamma_{cs}(G)=2$;
		}
	\ElseIf{$(|X| =3$ or $|Y|=3)$}{
		$\gamma_{cs}(G)=3$;
		}
	\ElseIf{$(|X| \geq 4$ and $|Y|\geq 4)$}{
		$\gamma_{cs}(G)=4$;
		}
	}
}
	
\If{$(k \geq 3)$}{
	\If{$(|Y_1| >1$ and $|X_k|>1)$}{
		$\gamma_{cs}(G)=4$;
		}
 	\ElseIf{$((|X_1| >1$ and $|Y_1|=1)$ and $(|Y_k| >1$ and $|X_k|=1))$}{
 		Let $G'=G[ \cup_{i=2}^{k-1} (X_i \cup Y_i)]$;\\
		$\gamma_{cs}(G)=|X_1|+ |Y_k|+$CSDN$\_$Chain$(G',k-2)$;
		}
	\ElseIf{$(|X_1| >1$ and $|Y_1|=1)$}{
 		Let $G'=G[ \cup_{i=2}^k (X_i \cup Y_i)]$;\\
		$\gamma_{cs}(G)=|X_1|+$CSDN$\_$Chain$(G',k-1)$;
		}
	\ElseIf{$(|Y_k| >1$ and $|X_k|=1)$}{
 		Let $G'=G[ \cup_{i=1}^{k-1} (X_i \cup Y_i)]$;\\
		$\gamma_{cs}(G)=|Y_k|+$CSDN$\_$Chain$(G',k-1)$;
		}
	\Else{
		\If{$(|X|=3$ or $|Y|=3)$}{
 			$\gamma_{cs}(G)=3$;
			}
		\Else{
			$\gamma_{cs}(G)=4$;
			}
		}
}
\Return	$\gamma_{cs}(G)$;}
\caption{\textbf{CSDN$\_$Chain$(G,k)$}}
\end{algorithm}

\begin{thm}\label{csdnchain}
Given a connected chain graph $G=(X,Y,E)$ with proper ordered chain partition $X_1,X_2, \ldots, X_k$ and $Y_1,Y_2, \ldots, Y_k$ of $X$ and $Y$. Then, the cosecure domination number of $G$ can be computed in  polynomial-time.
\end{thm}

\newpage

\section{Conclusion}
We resolved the complexity status of the \textsc{Minimum Cosecure Domination} problem on various important graph classes, namely, chain graphs, chordal bipartite graphs, star-convex bipartite graphs, comb-convex bipartite graphs and bounded tree-width graphs. It was known that the \textsc{Cosecure Domination Decision} problem is NP-complete for bipartite graphs. Extending this, we showed that the problem remains NP-complete even when restriced to star-convex bipartite graphs, comb-convex bipartite graphs and chordal bipartite graphs, which are all subclasses of bipartite graphs. Further, we have proved that the problem is NP-complete for doubly chordal graphs. On the positive side, we proved that the \textsc{Minimum Cosecure Domination} problem is efficiently solvable for chain graphs and bounded tree-width graphs. Naturally, it would be interesting to do the complexity study of the \textsc{Minimum Cosecure Domination} problem in many other important graphs classes for which the problem is still open.

%
%
%
%
%
%
%
%

%

%
%
%
%
%
\end{document}